\documentclass{appolb}
\usepackage[utf8]{inputenc}
\usepackage{epsf}
\usepackage{latexsym,amssymb,euscript}
\usepackage[dvips]{graphicx}
\usepackage[numbers,sort&compress]{natbib}
\usepackage{amsmath}
\usepackage{nicefrac}
\usepackage{slashed}
\usepackage{booktabs}
\usepackage{hyperref}
\usepackage{braket}
\usepackage{chngcntr}
\usepackage{bbold}
\usepackage{graphics}
\usepackage{graphicx}
\usepackage{mciteplus}
\usepackage{caption}
\usepackage{subcaption}
\usepackage{pdfpages}
\usepackage[titletoc]{appendix}
\graphicspath{{./figures/}}
\hypersetup{
 linktocpage = true,
 urlcolor = black,
 colorlinks = true,
 linkcolor = urlblue,
 anchorcolor = urlblue,
 citecolor = urlblue,
 pdfstartview = {XYZ null null 1.25} 
           }
\usepackage{pstricks}
\usepackage{color}
\usepackage{xcolor}
\definecolor{urlblue}{rgb}{0.2,0.4,0.7}
\definecolor{citegreen}{rgb}{0,0.4,0.2}
\definecolor{linkred}{rgb}{0.9,0.2,0.1}
\usepackage{float}
\definecolor{orcidlogocol}{HTML}{A6CE39}
\begin{document}
\title{The QCD shockwave approach at NLO: towards precision physics in gluonic saturation
\thanks{Presented at “Diffraction and Low-$x$ 2022”, Corigliano Calabro (Italy), September
24-30, 2022.}%
}
\author{Samuel Wallon, 
\address{Universit\'e Paris-Saclay, CNRS/IN2P3, IJCLab, 91405, Orsay, France}
}

\maketitle
\begin{abstract}
We review the recent developments of the QCD shockwave approach at Next-to-Leading Order. The general method of this effective action is sketched, and illustrated by the case of diffractive processes.
\end{abstract}
  
\section{Introduction}

At very high energies, one of the most intriguing phenomena of strong interaction is the existence of collective effects in nucleons and nuclei (target), named gluonic saturation. This is due to the dominance of particular Fock states made of a large number of weakly coupled gluons, which should recombine together, taming the growth of their distribution. Interestingly, there exists a window for which such a dynamics can be accessed based on perturbative methods: for any hard process, e.g. Deep Inelastic Scattering (DIS) governed by the virtuality   $Q^2$ of the photon probe and momentum fraction $x$, gluonic saturation should have an impact on observables, with a perturbative control, whenever $\Lambda_{QCD}^2 \ll Q^2 \lesssim Q_s^2$, where the saturation scale $Q_s^2 \sim (A/x)^{1/3}$. The key feature of $Q_s^2$ is that it increases with the center-of-mass energy (i.e. decreasing $x$) as well as with the nuclear mass number $A.$ This explains why LHC, in ultra-peripheral $Ap$ or $AA$ high-energy collisions, using a heavy ion as a source of photons to probe $p$ or $A$, or EIC with real or virtual photons used as probes of heavy nuclei, could give access to gluonic saturation in this perturbative window, much wider than at HERA. Still, a clear evidence of saturation remains to be exhibited experimentally without any doubt, and the dynamics of this collective state should be characterized, since it plays a key role as the initial state in $AA$ collisions leading to the production of quark-gluon plasma. This amounts to reach precision physics, namely controlling both the dynamics of the collective gluonic field emitted from the target as well as the way a given probe couples to this target,
with a precision at the level of Next-to-Leading Order (NLO).

\section{The framework}

Generically, either at the level of the amplitude for a diffractive process, or at the level of the cross-section for an inclusive process, at high-energy (large $s$) the 
color singlet $C=+1$ state exchanged in $t-$channel is a "Pomeron".
Restricting to the non saturated dynamics, one might describe 
this Pomeron in the linear Balitsky-Fadin-Kuraev-Lipatov (BFKL) regime~\cite{Fadin:1975cb, Kuraev:1976ge, Kuraev:1977fs, Balitsky:1978ic}, with next-to-leading logarithmic (NLL) precision~\cite{Fadin:1998py,Ciafaloni:1998gs,Fadin:2004zq,Fadin:2005zj} \footnote{For a recent review on tests of BFKL through semi-hard processes involving jets and hadrons, see Ref.~\cite{Celiberto:2020wpk}.}.
Basically, BFKL resums logarithmic contributions of type
\begin{equation}
\sum_n (\alpha_s \, \ln s)^n \ + \ \alpha_s \, \sum_n (\alpha_s \, \ln s)^n \, + \, \cdots
\end{equation}
in which the first kind of terms are leading logarithmic (LL) while the second are next-to-leading logarithmic (NLL).

At even higher energies, beyond these resummations, when multiple interactions between the probe and the target can occur due to the fact that the target is made of gluons with very high occupation numbers, 
one should additionally resum\footnote{A $\gamma$ or $\gamma^*$ probe looks like a $q \bar{q}$ dipole during the high-energy process;  $\alpha_s^2 A^{1/3}$ is the typical 
order of magnitude of the scattering amplitude through 2-gluon exchange between this dipole probe and a dipole inside a large nucleus $A$.} powers  of $\alpha_s^2 A^{1/3}$. Through this dynamics, an underlying system of packed gluons emerges. This new state of matter which dominates the tower of Fock states is named
color glass condensate (CGC).
The
 Pomeron is now either built from  Balitsky's high energy operator expansion~\cite{Balitsky:1995ub, Balitsky:1998kc, Balitsky:1998ya, Balitsky:2001re}, 
or from the hamiltonian  formulation~\cite{JalilianMarian:1997jx,JalilianMarian:1997gr,JalilianMarian:1997dw,JalilianMarian:1998cb,Kovner:2000pt,Weigert:2000gi,Iancu:2000hn,Iancu:2001ad,Ferreiro:2001qy}, which satisfies the Balitsky-Jalilian Marian-Iancu-McLerran-Weigert-Leonidov-Kovner (B-JIMWLK) evolution equation. We will now briefly sketch the first approach. 

Let us introduce a light-cone basis composed of $n_1 $ and $n_2$, with $n_1 \cdot n_2 = 1$, defining the $+/-$ direction respectively.
Consider a reference frame, named projectile frame, such that the target moves ultra-relativistically and such that 
$s = (p_\gamma + p_0)^2 \sim 2 p_\gamma^+ p_0^- \gg \Lambda_{\text{QCD}}^2$, $s$ also being larger than any other scale.  Particles on the projectile side are moving in the $n_1$ (i.e. $+$) direction while particles on the target side have a large component along $n_2$ (i.e. $-$ direction).

\begin{figure}[h]
%\psfrag{Y}{$\!\!\eta$}
\begin{center}
\includegraphics[width=7cm]{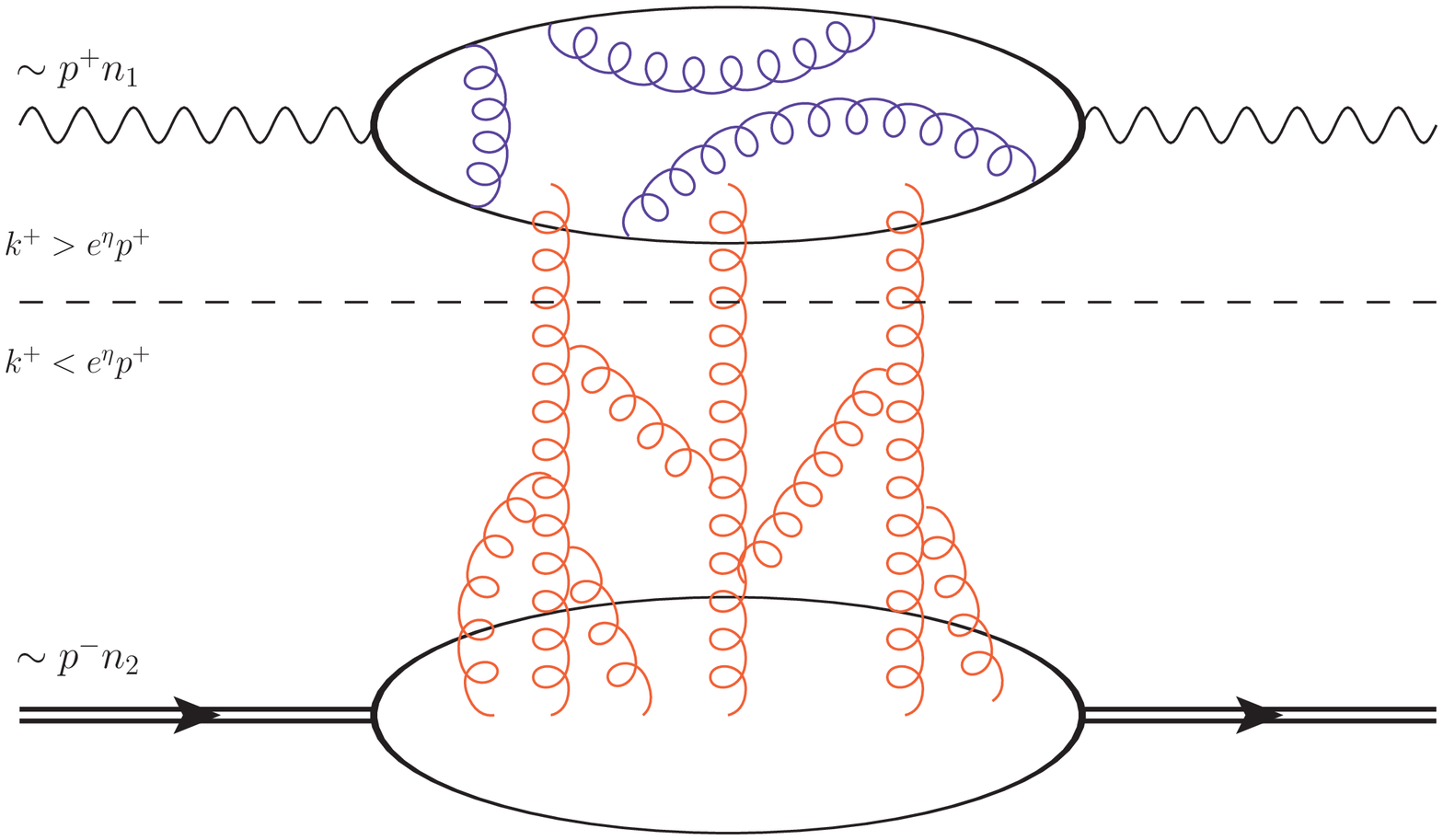}
\end{center}
\caption{Separation between classical (``slow'' gluons) and quantum (``fast'' gluons) modes.}
\label{Fig:separation}
\end{figure}

The shockwave formalism is an effective approach to deal with gluonic saturation. The gluonic field between the probe and the target $A$ is separated into external background fields $b$ (resp. internal fields $\mathcal{A}$) depending on whether their $+$-momentum is below (resp. above) the arbitrary rapidity cut-off $e^\eta p_\gamma^+$, with $\eta < 0$,
see fig.~\ref{Fig:separation}. This corresponds to splitting the gluonic field between "{{fast}}" and "{{slow}}" gluons,
\begin{eqnarray}
\mathcal{A}^{\mu a} (k^+, k^- ,\vec{k}\,) & = & {{A_\eta^{\mu a}\,(|k^+| > e^{\eta}p^+,k^-,\vec{k}\, )}} \qquad \hbox{{quantum part}} \nonumber \\
& + & {{b_\eta^{\mu a}(|k^+| < e^{\eta}p^+,k^-,\vec{k}\, )}}\qquad \hbox{{\ \ classical part.}} 
\end{eqnarray}

\begin{figure}[h]
\begin{center}
\includegraphics[width=5.5cm]{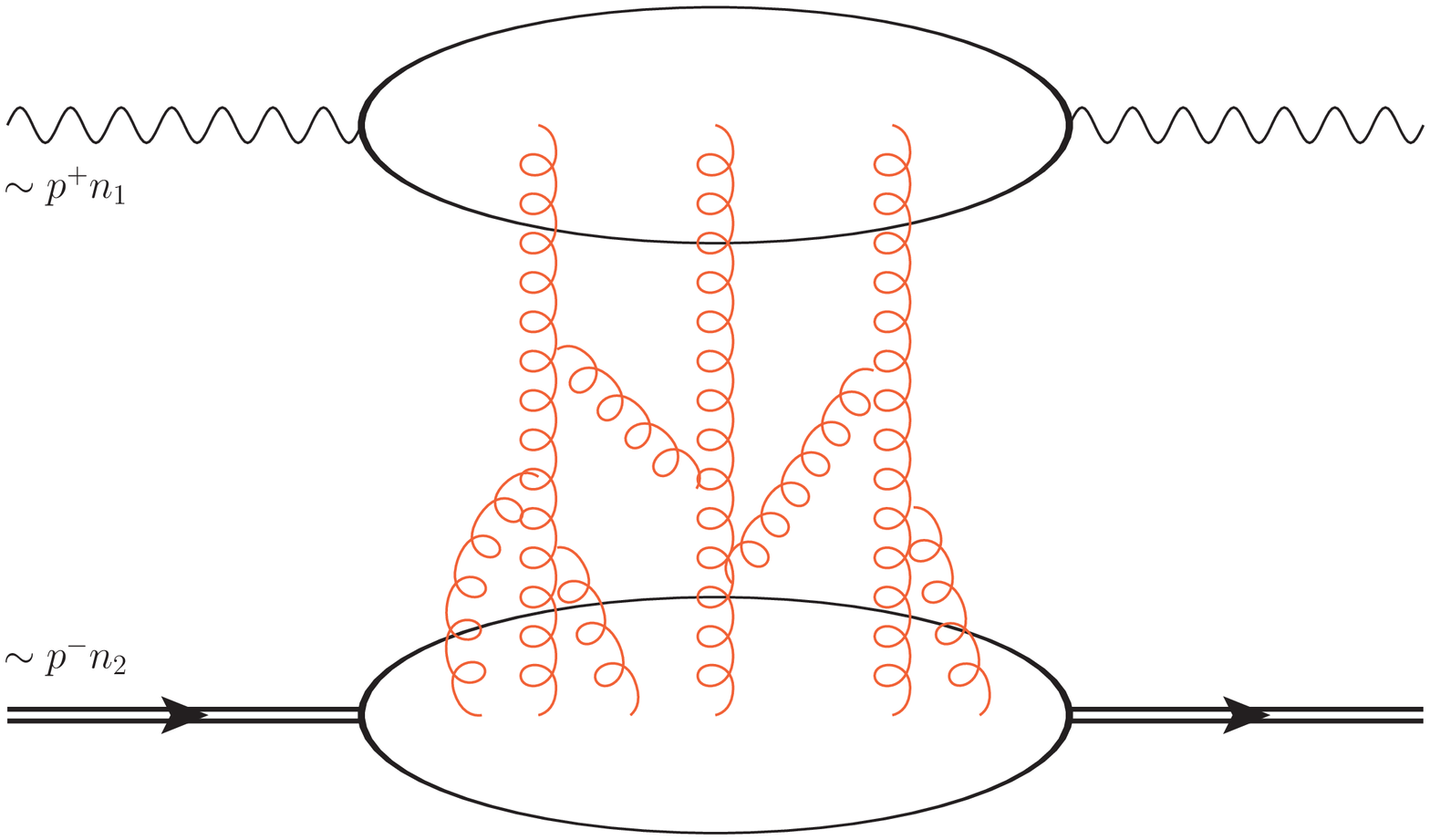}
\raisebox{1.3cm}{$\xrightarrow[]{\text{{boost}}}$}
\includegraphics[width=5.9cm]{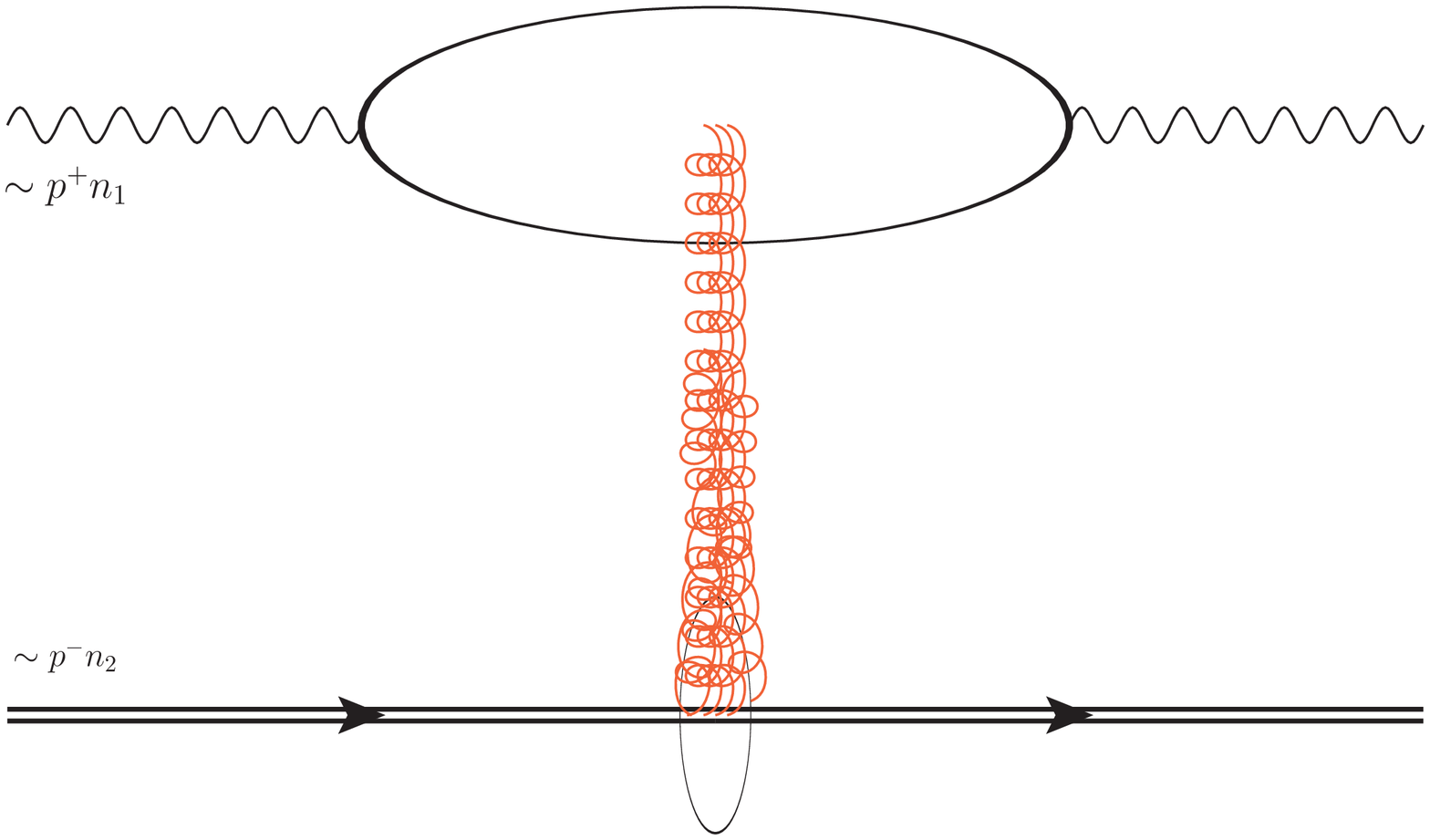}
\end{center}
\caption{Boosting from the target frame to the projectile frame.}
\label{Fig:boost}
\end{figure}

Moving from the target frame to the projectile frame, 
with a large longitudinal {boost} $\Lambda \propto \sqrt{s}$, the external fields take the shockwave form, see fig.~\ref{Fig:boost},  
\begin{equation}
    b^\mu (x) = b^-(x_\perp) \delta (x^+) n_2^\mu \,.
\end{equation}

Using the light-cone gauge $n_2 \cdot A = $ is particularly suitable with this shockwave structure. Indeed, this implies that $b \cdot A=0$, leading to simple Feynman rules in this effective field theory, when starting from the QCD lagrangian.

The resummation of all order interactions with this external field, i.e. the propagator in the shockwave background field, is a high-energy Wilson line, located exactly at $x^+ =0$:
\begin{equation}
    U_{\vec{z}} = \mathcal{P} \exp \left(i g \int d z^+ b^-(z)\right)\,,
\end{equation}
where $\mathcal{P}$ is the usual path ordering operator.

The small-$x$ factorization applies here and the scattering amplitude is the convolution of the projectile impact factor and the non-perturbative matrix element of operators from the Wilson lines operators on the target states. 
One of such operators is the dipole operator, which in the fundamental representation of $SU(N_c)$ takes the form
\begin{equation}
\!\!\left[\operatorname{Tr} \!\left(U_1 U_2^\dag\right)-N_c\right]\!(\vec{p_1},\vec{p}_2)\! = \!\!\int \! d^d \vec{z}_{1} d^d \vec{z}_{2\perp} e^{- i \vec{p}_1 \cdot \vec{z}_1} e^{- i \vec{p}_2 \cdot \vec{z}_2} \!\left[\operatorname{Tr} \left(U_{\vec{z}_1} U_{\vec{z}_2}^\dag\right)-N_c\right],
\end{equation}
where
$\vec{z}_{1,2}$ are the transverse positions of the $q,\bar{q}$ coming from the photon and $\vec{p}_{1,2}$ their respective transverse momentums kicks from the shockwave.
The factorized scattering amplitude then reads
\begin{equation} \nonumber
 \mathcal{A}^\eta 
 = \int \!  d^{D-2} \vec{z}_1 d^{D-2} \vec{z}_2 \, {{\Phi^\eta(\vec{z}_1,\vec{z}_2\,)}} \, {{\langle P^\prime | [\mathrm{Tr}(U_{\vec{z}_1}^\eta U_{\vec{z}_2}^{\eta\dagger})-N_c ] | P 
 \rangle}}.
\end{equation}

Generally speaking, the Wilson-line operators evolve 
with respect to rapidity according to the Balitsky hierarchy.
Restricting to the case of a dipole operator, it reduces to the Balitsky-Kovchegov (BK) 
equation~\cite{Balitsky:1995ub, Balitsky:1998kc, Balitsky:1998ya, Balitsky:2001re, Kovchegov:1999yj, Kovchegov:1999ua} in the large $N_c$ limit.

The NLO evolution kernel are now known both in the shockwave framework~\cite{Balitsky:2008zza,Balitsky:2013fea,Grabovsky:2013mba,Balitsky:2014mca} and in the hamiltonian approach~\cite{Kovner:2013ona,Caron-Huot:2015bja,Lublinsky:2016meo}.
Besides, the number of processes for which the impact factors are known with a NLO accuracy is rather small. For inclusive processes, this list contains
 inclusive DIS~\cite{Balitsky:2010ze,Balitsky:2012bs,Beuf:2022ndu}, inclusive photoproduction of dijets~\cite{Altinoluk:2020qet,Taels:2022tza},  
photon-dijet production in DIS~\cite{Roy:2019hwr}, dijets in DIS~\cite{Caucal:2021ent,Caucal:2022ulg}, single hadron~\cite{Bergabo:2022zhe}  and dihadrons production in DIS~\cite{Bergabo:2022tcu,Iancu:2022gpw}. For diffractive processes, NLO impact factors are known for exclusive dijets~\cite{Boussarie:2014lxa,Boussarie:2016ogo,Boussarie:2019ero} and exclusive light meson production~\cite{Boussarie:2016bkq,Mantysaari:2022bsp}, exclusive quarkonium production~\cite{Mantysaari:2021ryb,Mantysaari:2022kdm}, and inclusive diffractive DIS~\cite{Beuf:2022kyp}. Recently, the 
impact factor for diffractive production of a pair of hadrons at large $p_T$, in $\gamma^{(*)}$ nucleon/nucleus scattering was obtained, in the most general kinematics~\cite{Fucilla:2022wcg}.

\section{Application to diffractive processes}

Diffractive processes have been revealed experimentally by H1 and ZEUS experiments at HERA, for the first time in the semi-hard regime in which a hard scale allows one to describe such processes from first principles, relying on QCD. Due to the large center-of-mass energy, these processes provide an access to the regime of very high gluon densities~\cite{Wusthoff:1999cr,Wolf:2009jm}.
HERA showed that almost 10~\%  of the $\gamma^* p \to X$ DIS events present
 a rapidity gap between the proton remnants 
and the hadrons 
coming from the fragmentation region of the initial virtual photon. These
events are called diffractive deep inelastic scattering (DDIS), and look like
$\gamma^* p \to X  \, Y$~\cite{Aktas:2006hx,Aktas:2006hy,Chekanov:2004hy,Chekanov:2005vv,Aaron:2010aa,Aaron:2012ad,Chekanov:2008fh,Aaron:2012hua}, $Y$ being either the outgoing proton or one of its low-mass excited states, and $X$ the diffractive final state. One can further select specific diffractive states, like jet(s), a single meson, or a pair of hadrons.
The existence of a rapidity gap between $X$ and $Y$ leads naturally to describe diffraction through a Pomeron exchange in the $t-$channel between these $X$ and $Y$ states. 

In the collinear framework, a QCD factorization theorem~\cite{Collins:1997sr} justified by the existence of a hard scale, the photon virtuality $Q^2$ of DIS, allows to describe the scattering amplitude as a  convolution of
a coefficient function with diffractive parton distributions, the latter describing the partonic content of the Pomeron. At high energies, it is natural to model the diffractive events by a {\em direct} Pomeron contribution involving the coupling of a Pomeron with the diffractive state $X$ of invariant mass $M.$ In the very high energy regime, the $t-$channel state can be seen as a color-singlet QCD shockwave.

\begin{figure}[h]
\vspace{0.02cm}
\scalebox{1.1}{\begin{picture}(100,110)
\put(75,0){\includegraphics[width=6.5cm]{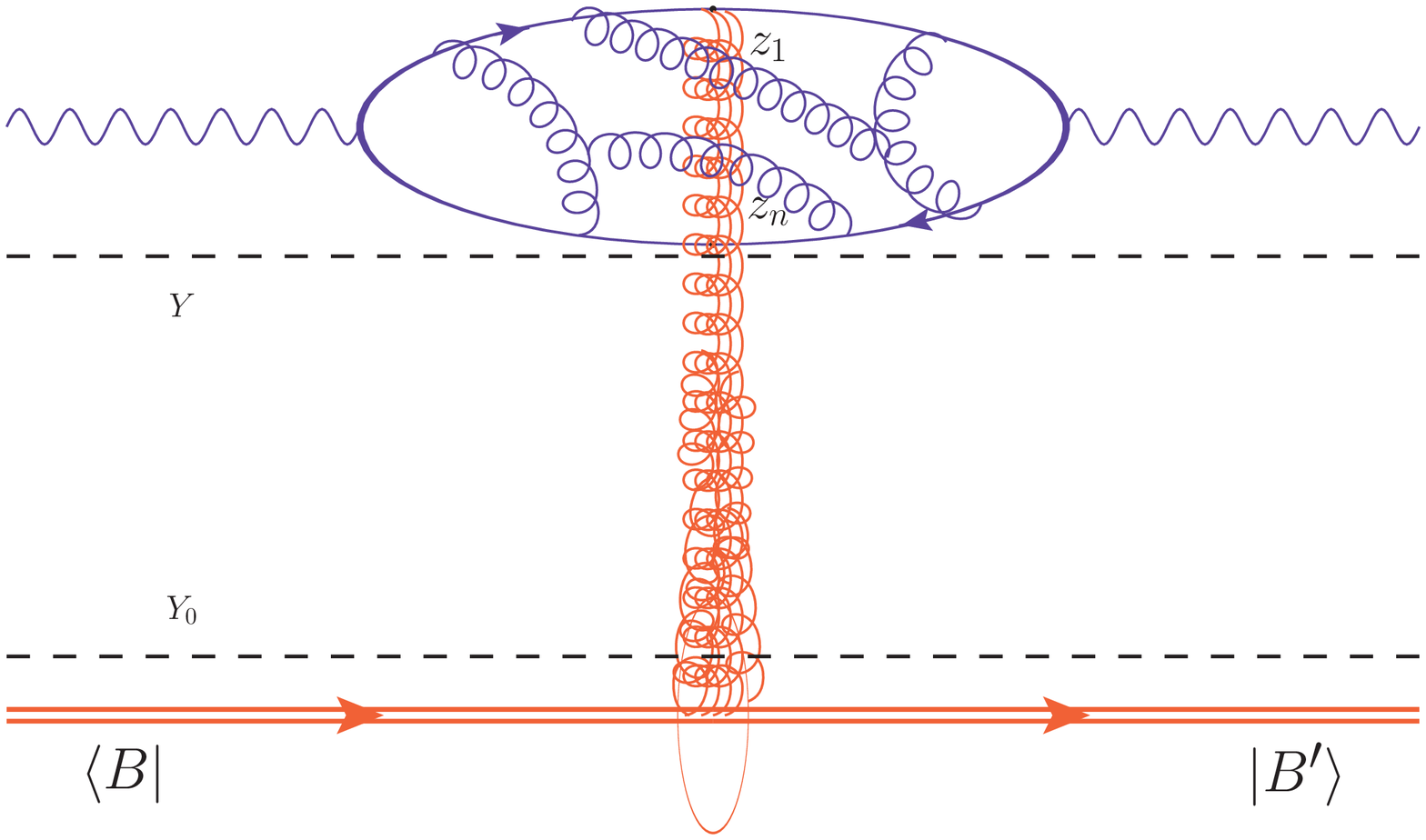}}
\put(205,85){\includegraphics[width=2cm]{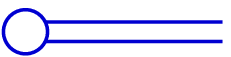}}
\end{picture}}
\vspace{.2cm}
\caption{The different building blocks involved in the shockwave description of the scattering amplitude of diffractive exclusive production of a meson. The NLO impact factor corresponds to the case of one single (blue) gluon in the upper part.}
\label{fig:diffractive-process}
\end{figure}

The factorized amplitude of a diffractive process 
\begin{eqnarray}
{A} = \int\!\!  d\vec{z}_1 ... d\vec{z}_n \,\, {{\Phi(\vec{z}_1,...,\vec{z}_n)}} 
\,{{\langle P^\prime | U_{\vec{z}_1}...U_{\vec{z}_n} | P \rangle}} \nonumber
\end{eqnarray}
is illustrated in fig.~\ref{fig:diffractive-process}. Its evaluation within a NLO precision amounts to compute the upper impact factor using the effective Feynman rules, build non-perturbative models for the matrix elements of the Wilson line operators acting on the target states, 
solve the B-JIMWLK evolution for these matrix elements with such non-perturbative initial conditions at a typical target rapidity $Y_0$, and  
evaluate the solution at a typical projectile rapidity $Y$, or at the rapidity of the slowest gluon, and 
convolute the solution and the impact factor. A preliminary example, in the case of forward exclusive diffractive dijet applied to HERA data, can be found in~\cite{Boussarie:2019ero}.

\section{Summary and outlook}

A new era in the study of gluonic saturation has started, with the goal of reaching a NLO precision and thus to reveal and describe the CGC, an intriguing new state of matter, made of collective weakly coupled gluonic excitations. Among a large collection of accessible processes, exclusive diffraction is particularly appealing since    
it allows one to probe the impact parameter dependence of the non-perturbative scattering amplitude, and furthermore, to access the 5-dimensional Wigner function at low-$x$.

\end{document}